\documentclass[twocolumn,prl,preprintnumbers,amsmath,amssymb,superscriptaddress]{revtex4}

\usepackage{dcolumn}
\usepackage{bm}
\usepackage[dvips]{pstcol,graphicx}
\usepackage{pstricks}

\begin{document}

\title{Magnetic order and valency at La$_{0.7}$Sr$_{0.3}$MnO$_{3}$/SrTiO$_{3}$ interfaces}

\author{H. Zenia}
 \email{H.Zenia@Sheffield.ac.uk}
 \affiliation{Department of Physics and Astronomy, University of Sheffield, Sheffield, S3 7RH, UK }
 \affiliation{ Daresbury Laboratory, Daresbury, Warrington WA4 4AD, UK }
\author{G. A. Gehring}
 \affiliation{Department of Physics and Astronomy, University of Sheffield, Sheffield, S3 7RH, UK }
\author{W. M. Temmerman}
 \affiliation{ Daresbury Laboratory, Daresbury, Warrington WA4 4AD, UK }

\begin{abstract}
We report on first principles calculations of the electronic structure of
La$_{0.7}$Sr$_{0.3}$MnO$_{3}$/SrTiO$_{3}$ junction with two possible types of
interface terminations. We find that the La$_{0.7}$Sr$_{0.3}$O/TiO$_{2}$
interface preserves the interlayer ferromagnetic coupling between the
interface MnO$_{2}$ layer and the bulk. The other interface, MnO$_{2}$/SrO,
favours antiferromagnetic coupling with the bulk. By inserting two unit cells
of undoped LaMnO$_{3}$ at the interface the ferromagnetism is recovered. This
is understood in terms of the doping level and the mobility of carriers near
the interface.

\end{abstract}


\maketitle

Spintronic devices use the information carried by the spin of
electrons as well as their charge, and spin dependent tunnelling
lies at the heart of their operation. They depend on having a
strongly polarised or half metallic ferromagnet from which
polarised carriers can tunnel. The rare earth manganites,
particularly La$_{0.7}$Sr$_{0.3}$MnO$_{3}$ (LSMO), are good
candidates for tunnel devices because there is evidence that they
have a high polarisation $P\sim1$ \cite{park98,nadgorny01}. There
are three factors that should be considered for a good tunnelling
device.  First the magnetism of the surface layer should not be
much lower than the bulk, second the electronic spin polarisation,
$P$, at the surface should be high, and third if $P<1$ the
velocities of the majority carriers should be high and that of the
minority carriers should be low \cite{mazin99}. A perfect
tunneling magnetoresistance (TMR) would be obtained if the
minority spin states at the interface were actually localised.

Insulating SrTiO$_{3}$ (STO) is one of the most promising
materials to use as a tunnel barrier. It is a good lattice match
to LSMO and also has a small band gap so that the tunnelling rates
are high. Hence the importance of understanding the nature of the
interface between these two perovskites. The usual surface of STO is a
TiO$_{2}$ layer \cite{izumi98}. If a film of LSMO is deposited on top, it will
normally contain an equal number of La$_{0.7}$Sr$_{0.3}$O and MnO$_{2}$ layers
and so will be terminated by a MnO$_{2}$ layer \cite{izumi98}. A tunnel
barrier of STO will thus start with a SrO layer and finish with a TiO$_{2}$
layer so that junctions(interfaces) of both types occur. 

There have recently been studies
\cite{samet03,bowen03,bowenthesis,yamada04,izumi01,izumi02} of the magnetic 
properties of LSMO/STO interface when it is grown as
La$_{0.7}$Sr$_{0.3}$O/TiO$_{2}$  and as MnO$_{2}$/SrO. It is convenient to
refer to these as TiO$_{2}$ and SrO interfaces respectively. Bowen \textit{et
  al.} \cite{bowen03,bowenthesis} grew tunnel structures in which both
interfaces were of the TiO$_{2}$ type and obtained a value of the  tunnelling
magnetoresistance (TMR) of 1800\%. Yamada \textit{et al.} \cite{yamada04}
showed that the magnetism and TMR at the SrO interface were enhanced when two
layers of undoped LaO were grown adjacent to the SrO layer, however even then
the magnitude of the observed TMR (170\%) is far below what was observed for
the TiO$_{2}$ interface.

In this letter we give the first microscopic analysis of the nature of the
magnetism for LSMO at a TiO$_{2}$ and at a SrO interface, when zero, one and
two layers of undoped LMO are grown adjacent to the interface . We address the
three important issues: the magnitude of the magnetic exchange between the
surface layer and the bulk, the density of states at the Fermi level both in
the bulk and at the surface and the value of the spin polarisation, $P$, at
the surface and finally the relative mobility of the majority and minority
carriers at the surface.

We address all the terminations in a self-consistent Self-Interaction
Corrected LSD (SIC-LSD) \cite{temmerman} calculation by means of sufficiently
large supercells (8-10 unit cells of LSMO and 8 unit cells of STO). The
SIC-LSD method has been  applied successfully to a variety of problems where
competition occurs between localisation and delocalisation of electrons in
``strongly'' correlated systems such as transition metal oxides
\cite{szotek93}.

In the present study we have used the experimental lattice parameter of
SrTiO$_{3}$, 7.38 a.u, for the whole system which has a cubic perovskite
structure. To model the LSMO system we use a  virtual La atom of atomic number
$Z-x$, where $x=0.3$ is the doping level. This approximation is very
reasonable in this case which has been confirmed by comparison to
supercell calculations \cite{banach}.

\begin{figure} 
\begin{pspicture}(-5cm,1cm)(3cm,7cm) 
\rput(1.9,3.765) {\includegraphics[height=6.4cm,width=2.2cm,clip]{figures/dos_bulk_2.eps}}
\rput(-2,4){\includegraphics[height=6cm,width=6cm,angle=-90,clip]{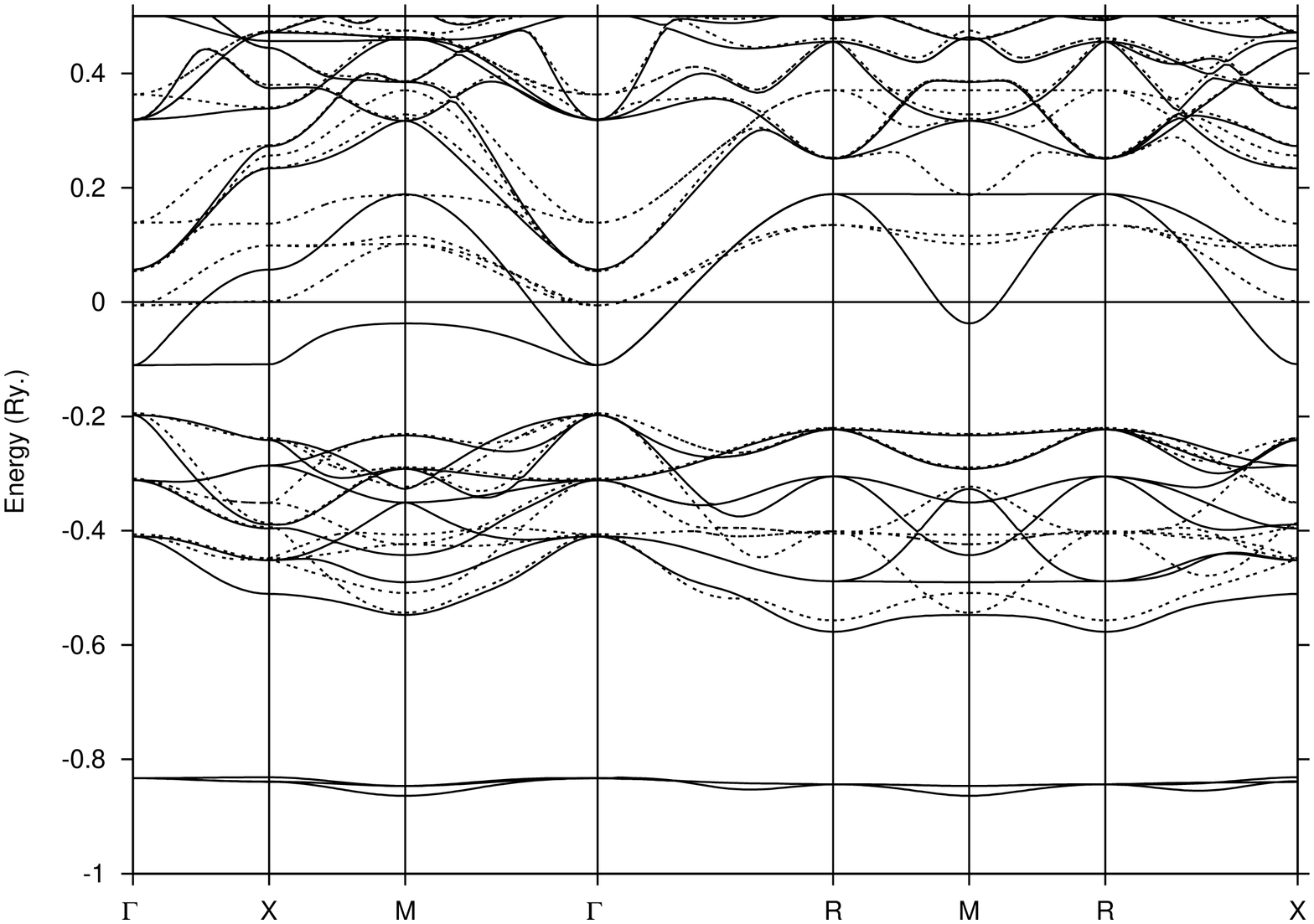}}
\end{pspicture}
\caption{\label{fig1} SIC-LSD band structure and density of states (DOS) of bulk LSMO
  in the ferromagnetic ground state with the three $t_{2g}$ orbitals
  localised. Full line is used for the majority  and the dotted line is for
  the minority spins .}

\end{figure}

We first apply our method to calculating the electronic structure for bulk LSMO
and STO. We find LSMO to be nearly half-metallic with the  Fermi level
($E_{F}$) lying at the bottom of the minority conduction band as shown on
Fig. \ref{fig1}. In the ground state the three $t_{2g}$ orbits are localised
and the total magnetic moment in the unit cell is of 3.47$\mu_{B}$
(experimental moment is $\sim 3.60\mu_{B}$). Note that in a half-metal the
total moment should be 3.70$\mu_{B}$. There has been much debate on whether
the manganites such as LSMO are really half-metals. Spin-polarised
photoemission spectroscopy measurements on LSMO showed a 100\% \cite{park98}
spin polarisation at $E_{F}$ while Andreev reflection experiment
\cite{nadgorny01} found existence of minority states at $E_{F}$. Early LSD
calculations found the system to be nearly half-metallic \cite{livesay99} and
recently using SIC and allowing for mixed valency this material has been
established to be half-metallic \cite{banach}. It was  argued \cite{pickett96}
that in manganites, like LSMO, the minority electrons localise because of the
random distribution of the trivalent and divalent ions  as well as their
narrow occupied bandwidth (Fig. \ref{fig1}). Consequently, the system behaves
as a transport half-metal \cite{mazin99}.

Bulk STO is known to be a band insulator and in our LDA
calculation we find that it has a gap of 2.09eV, smaller than its experimental
value of 3.25eV. The LDA is well-known to underestimate band-gaps and one
way to remedy this deficiency would be to use a GW \cite{godby88}
approximation. There is no SI correction in this system because
there are no localised electrons, the Ti ion being in the $d^{0}$
configuration.

We have considered both interface terminations (TiO$_{2}$ and SrO) by
performing two separate calculations with symmetric supercells so that in each
case we have only one kind of termination present. Using the following
notation: L for La$_{0.7}$Sr$_{0.3}$O, M for MnO$_{2}$, T for TiO$_{2}$ and S
for SrO, the first supercell is LM[LM]$_{6}$LT[ST]$_{6}$ST which has two
TiO$_{2}$-type interfaces. The second is M[LM]$_{6}$LMST[ST]$_{6}$S which  has
two SrO-type interfaces. The subscript is for the number of formula units. We
also consider the cases where one and two La$_{0.7}$Sr$_{0.3}$O layers at the
SrO interface are replaced by undoped LaO . Within the layers we have assumed
ferromagnetic (FM) coupling and therefore used one Mn atom per layer. This is
a very reasonable assumption since both $e_{g}$  ($d_{3z^{2}-r^{2}}$ and 
$d_{x{^2}-y^{2}}$) orbits have significant hopping integrals
within the layers and it is known from the double-exchange model
that the kinetic energy mediates the FM coupling. Between the
layers, on the other hand, only the
$d_{3z^{2}-r^{2}}$-$d_{3z^{2}-r^{2}}$ hopping integral is nonzero,
and this has significant influence on the interlayer magnetic
coupling \cite{zenia}. Bulk LSMO is ferromagnetic so we
investigate the magnitude of the exchange interaction between the
surface layer and the bulk by considering the energy difference
$E_{AF}-E_{FM}$ of the supercell between when the  interface
MnO$_{2}$ layer is ferromagnetically (FM) and
antiferromagnetically (AF) aligned with the bulk  and compare this
to the bulk value. Experimentally, this is corresponds to the
differences observed between the interface and bulk Curie
temperatures ($T_{C}$) \cite{garcia04}. We find that the exchange
estimated in this way depends on the interface.
\begin{table}[h!]
\centering
\caption{\label{tab0} Exchange energies and magnetic moments in the bulk and
  at TiO$_{2}$, SrO  and SrO + 1(2) LaO interfaces. The energies are
given in mRy and the moments in units of $\mu_{B}$.}

\begin{tabular}{|l|c|c|c|c|c|}
\hline
                     &Bulk   &TiO$_{2}$   &SrO    &SrO + 1 LaO &SrO + 2 LaO\\
\hline
\hline
$E_{AF}-E_{FM}$      &6.0    &3.5         &-6.3   &-5.5       &3.0\\
\hline
Moment              &3.47    &3.21        &3.31   &3.34       &3.31\\
\hline
\end{tabular}
\end{table}

The values of the exchange energies are given in the Table
\ref{tab0}. In the case of TiO$_{2}$ interfaces we find a value of
the surface exchange $E_{AF}-E_{FM}$ which is 58\% of the bulk
value indicating that the surface magnetism is well coupled to the
bulk.  However for SrO interfaces we find that $E_{AF}-E_{FM}$ is
negative, indicating that the surface layer is coupled
antiferromagnetically to the bulk, this is clearly not a good
candidate for a large TMR. The magnitude of the antiferromagnetic
coupling is reduced if one undoped layer of LaO is inserted as the
penultimate layer to the interface. However the coupling becomes ferromagnetic
when two layers of undoped LaMnO are grown next to the surface.  This is the 
configuration that Yamada et al \cite{yamada04} found gave better
tunnelling magnetoresistance. In this case the value of
$E_{AF}-E_{FM}$ is 50\% that of the bulk indicating that the
surface magnetism is comparable to that of the TiO$_{2}$
interface.
\begin{table}[h!]
\centering
\caption{\label{tab1} Total energy differences for the different localisation scenarios
and magnetic coupling at TiO$_{2}$  and SrO  interfaces. The energy is
given in mRy per interface.}

\begin{tabular}{|l|c|c|c|}
\hline
  Coupling at interface    &SIC correction                  &TiO$_{2}$   &SrO \\
\hline
\hline
FM                         &$t_{2g}$                         &\textbf{0.0}      &\textbf{0.0}      \\
\hline
FM                         &$t_{2g}+(3z^{2}-r^{2})$          &12.9     &39.8     \\
\hline
FM                         &$t_{2g}+(x^{2}-y^{2})$           &23.5     &37.9     \\
\hline
\hline
AF                          &$t_{2g}$                        &\textbf{3.5}      &\textbf{-6.3}    \\
\hline
AF                          &$t_{2g}+(3z^{2}-r^{2})$         &8.9     &46.0     \\
\hline
AF                          &$t_{2g}+(x^{2}-y^{2})$          &22.4     &30.6     \\
\hline
\end{tabular}
\end{table}

The SIC calculations indicate that it is favourable if only the $t_{2g}$
states are localised.  However we can infer from the results presented in
Table \ref{tab1} that the energy required to localise a further $e_{g}$
electron depends on the symmetry of the electron state, the magnetisation and
the nature of the interface.  We have a qualitative understanding of these
effect from the fact that it is only the electrons in the $(3z^{2}-r^{2})$
orbits that transfer between layers. In the case of the TiO interface
localising this orbit will switch off some of the ferromagnetic coupling and
hence this has a higher penalty for the FM state than for the AF state, for
this interface there is a larger penalty to localise the $(x^{2}-y^{2})$ state
so indicating that in the conducting state the occupation of this state is
lower than for $(3z^{2}-r^{2})$ as was found to be favourable for
ferromagnetic coupling \cite{zenia}.  All the localisation energies for the
$e_{g}$ states are higher for the SrO interface than for the TiO$_{2}$
interface due to the lower electron density in the MnO$_{2}$ layer adjacent to
the the SrO interface.

\begin{figure} 
\includegraphics[scale=0.5,clip]{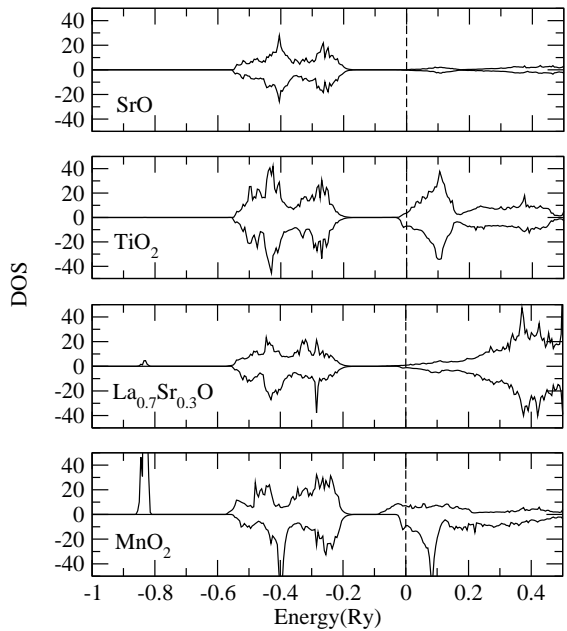}
\caption{Layer-projected DOS (states/layer/Ry) for the
  La$_{0.7}$Sr$_{0.3}$O/TiO$_{2}$ type of interface. Shown are the four
  layers making up the interface.} 
\label{fig2}
\end{figure}

\begin{figure} 
\includegraphics[scale=0.5,clip]{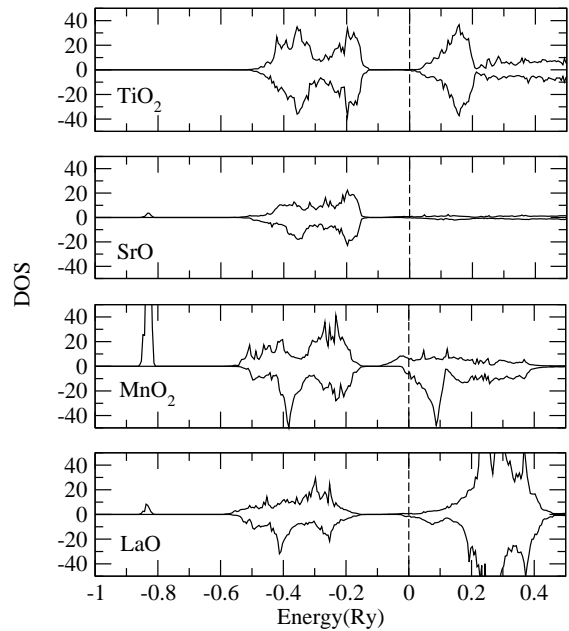}
\caption{Layer-projected DOS (states/layer/Ry) for the MnO$_{2}$/SrO type  of
  interface with two LaO layers substituted for La$_{0.7}$Sr$_{0.3}$O near the
  interface. Shown are the four layers making up the interface.} 
\label{fig3}
\end{figure}

Considering the two situations where ferromagnetism is found to be
stable at the interface, we looked at spin polarisation at $E_{F}$ as well as
at the degree of localisation in the two spin channels. The latter is obtained
from the shape of the DOS. The DOS are shown in Figs. \ref{fig2} and
\ref{fig3} for the TiO$_{2}$ and the SrO + 2LaO interfaces. We find that the
polarisation is actually  negative (-10\%) for the TiO$_{2}$
interface. However, the electrons in the minority spin channel are much more
localised than those in the majority spin channel as can be seen from the DOS of Fig
\ref{fig2}. The SrO interface with two LaO layers inserted has a small
positive polarisation (4\%) but again the minority electrons are more
localised than the majority ones (Fig \ref{fig3}). Moderate to high TMR
effects were found in LSMO-based junctions. This is a clear indication that
when $P<1$ the value of the TMR is not given by the DOS alone but one should
take account of the relative degree of delocalisation of the carriers in the
two spin channels \cite{mazin99}. It is not possible, though, to see the 
localisation of the carriers, due to disorder, from the DOS. The
latter is smooth through any localisation transition. However the
minority spin electrons are more likely to be below the mobility
edge than the majority electrons. The DOS show that the occupied
bandwidth of the minority spin electrons is much reduced compared
to that of the majority ones. We, therefore, expect the arguments
of Pickett and Singh \cite{pickett96} to hold in the current
situation. The DOS show averaged properties because of the
integration over both parallel $\vec{k}_{||}$ and perpendicular
$\vec{k}_{\perp}$ momenta. The localisation can be even stronger
when considering only $\vec{k}_{\perp}$ which is relevant to
tunnelling.

The change in the magnetic coupling at the interface is related to two key
ingredients : the doping level and the mobility (kinetic energy)
of the $e_g$ electrons. In order to compare the amount of $e_g$ electrons on
each MnO$_{2}$ layer we integrated the layer-projected DOS from the bottom of
the conduction band up to $E_{F}$ bearing in mind that these bands originate
from the Mn $e_g$ orbitals with a small hybridisation with O 2p
orbitals. These values are given in Table \ref{tab2} for the four scenarios
considered: TiO$_{2}$ interface, SrO interface and the two cases when one or
two La$_{0.7}$Sr$_{0.3}$O layers are substituted by LaO at the SrO
interface. The calculated charges are also compared with estimates from an
ionic picture. The latter are obtained by assuming a contribution of $(1-x)/2$
electron from each neighbouring La$_{1-x}$Sr$_{x}$O layer to a particular
MnO$_{2}$ layer. If we consider the SrO interface for instance then the ionic
value is $0.7/2+0/2=0.35$, given that the MnO$_{2}$ layer is sandwiched
between a La$_{0.7}$Sr$_{0.3}$O layer which contributes $0.7/2$ and a SrO
layer which contributes 0 electrons. As expected there are deviations from the
ionic values because of band formation and hybridisation. The largest of these
deviations occurs at the TiO$_{2}$ interface where we find 0.68 electrons
whereas from the ionic picture we have 1.05 electrons. In this case, however,
and as can be seen from Fig. \ref{fig2}, there are a few electrons (0.41) in
the conduction band of the interface TiO$_{2}$ layer. The effect of the number
of $e_{g}$ carriers on magnetic coupling is seen when comparing the charge on
the interface MnO$_{2}$ layer between the two types of interfaces. The charge
at the SrO interface (0.41) has been reduced below the bulk value (0.67). This
difference accounts for the observed AF coupling for the SrO interface whereas
the FM solution is stable for the TiO$_{2}$. All the other layers have
roughly the same charge. However, the effect of the number of carriers cannot
be separated completely from the effect of their mobility as we will discuss
below.

\begin{table}[h!]
\centering
\caption{\label{tab2} Electronic charge on MnO$_{2}$ layers in the bulk and
near the TiO$_{2}$ and SrO interfaces. SrO +1(2) LaO means 1(2) layers of LaO
inserted at the SrO interface. Indices C and I stand for the centre and the
interface respectively. Mn$_{I-1}$ is the layer next neighbour to the interface.}

\begin{tabular}{|l|c|c|}
\hline
  Interface and doping     &Calculated charge                &Ionic charge\\
\hline
\hline
TiO$_{2}$: Mn$_{C}$                 &0.64                             &0.70          \\
TiO$_{2}$: Mn$_{I-1}$               &0.65                             &0.70          \\
TiO$_{2}$: Mn$_{I}$               &0.68                             &1.05          \\
\hline
SrO: Mn$_{C}$                &0.67                             &0.70          \\
SrO: Mn$_{I-1}$              &0.62                             &0.70          \\
SrO: Mn$_{I}$              &0.41                             &0.35          \\
\hline
SrO +1 LaO: Mn$_{C}$         &0.65                             &0.70          \\
SrO +1 LaO: Mn$_{I-1}$       &0.71                             &0.85          \\
SrO +1 LaO: Mn$_{I}$       &0.54                             &0.50          \\
\hline
SrO +2 LaO: Mn$_{C}$         &0.65                             &0.70          \\
SrO +2 LaO: Mn$_{I-1}$       &0.75                             &1.00          \\
SrO +2 LaO: Mn$_{I}$       &0.50                             &0.50          \\
\hline
\end{tabular}
\end{table}

The effect of the mobility of the $e_g$ electrons can be seen by comparing the
number of carriers in $e_g$ bands in the bulk and at the TiO$_{2}$
interface. Although this value is indeed slightly larger at the interface,
we find that ferromagnetic exchange is smaller at the interface as compared to
the bulk. This can be attributed to the fact that the carriers at
the interface are confined to a 2D motion which, normally, favours
antiferromagnetism whereas in the bulk the interlayer hopping mediates a
stronger FM coupling. In the case of the SrO interface the charge on the
interface MnO$_2$ layer is of 0.41, much smaller than the bulk value of
0.67. This corresponds to a doping $x=0.65$ in the ionic  picture. With this
amount of doping, well above 0.5, even the bulk material is in a AF, and
possibly charge ordered, state. This high level of doping combined with the
confinement of electrons due to the interface leads to a very strong tendency
to AF coupling to the bulk as found from our total energy calculation (see
Table \ref{tab1}). Upon inserting one LaO layer the charge at the interface
MnO$_{2}$ increases from 0.41 to 0.54, its ionic value being 0.50 in this
case. This value is the theoretical ionic limit for the SrO interface. The
coupling is still AF though. By introducing a second LaO layer we see indeed
that the charge at the interface remains $\sim 0.50$. The excess charge goes
to the MnO$_{2}$ layers adjacent to the interface. On these layers the
electrons have more mobility and as a result lead to stronger FM coupling
between bulk and interface. We see then that increasing the number of carriers
at, but not limited to, the interface MnO$_{2}$ layer leads to a FM coupling
between interface and bulk and also to the preservation of transport
half-metalicity. Considering only the case of two layers of LaO inserted at
the SrO interface of a STO/La$_{0.6}$Sr$_{0.4}$MnO$_{3}$ junction, an increase
of TMR (50\% to 170\%) was reported \cite{yamada04}.

In summary, we have studied the two possible interface
terminations of a LSMO/STO/LSMO junction and found that the
TiO$_{2}$ interface preserves the ferromagnetism and the transport
half-metallicity, characteristic of bulk LSMO, both of which are
very important for a high TMR. For the SrO interface, on the other
hand, it is necessary to add carriers to the interface and also to
the neighbouring MnO$_{2}$ layers in order to recover these
properties. The number of carriers and their mobility are crucial
for enhancing ferromagnetism at the interface. This work has
therefore provided a consistent interpretation to two different
sets of experimental data on LSMO/STO interfaces. It is hoped that
it will serve as a guide to experimentalists in order to improve
the efficiency of manganite-based tunnel junctions.

\end{document}